\documentclass[aip, apl, reprint]{revtex4-1}

\usepackage{graphicx}  
\usepackage{amsmath} 
\usepackage{dcolumn}   
\usepackage{bm}        
\usepackage{amssymb}   
\usepackage{epstopdf}
\usepackage{xcolor} 
\usepackage[normalem]{ulem}

\def \lf {L_{\phi}}
\def \BS {Bi$_2$Se$_3$~}

\begin{document}

\title{Enhanced coherence and decoupled surface states in topological insulators through structural disorder}
\date{\today}

\author{Abhishek Banerjee}
\affiliation{ 
Department of Physics, Indian Institute of Science, Bangalore 560012, India
}%

\author{R Ganesan}
\affiliation{ 
Department of Physics, Indian Institute of Science, Bangalore 560012, India
}%

\author{P S Anil Kumar}
\affiliation{ 
Department of Physics, Indian Institute of Science, Bangalore 560012, India
}%

\begin{abstract} To harness the true potential of topological insulators as quantum materials for information processing, it is imperative to maximise topological surface state conduction, while simulateneously improving their quantum coherence. However, these goals have turned out to be contradictory. Surface dominated transport in topological insulators has been achieved primarily through compensation doping of bulk carriers that introduces tremendous {\it electronic} disorder and drastically deteriorates electronic coherence. In this work, we use structural disorder instead of electronic disorder to manipulate electrical properties of thin films of topological insulator Bi$_2$Se$_3$. We achieve decoupled surface state transport in our samples and observe significantly suppressed carrier dephasing rates in the coupled surface state regime. As the film thickness is decreased, the dephasing rate evolves from a linear to a super-linear temperature dependence. While the former is consistent with Nyquist electron-electron interactions, the latter leads to significantly enhanced coherence at low temperatures and is indicative of energy exchange due to frictional drag between the two surface states. Our work opens up the way to harness topological surface states, without being afflicted by the deleterious effects of compensation doping.    

\end{abstract}
\maketitle

Three dimensional topological insulators(TIs) have emerged as an important class of quantum materials characterized by insulating states in the bulk but topologically protected conducting states on the surface~\cite{TIreview1,TIreview2}. The recent realization of compensation doping of residual bulk carriers in the Bi$_{x}$Sb$_{2-x}$Te$_{y}$Se$_{3-y}$ (BSTS) class of TIs~\cite{dopedTI1, dopedTI2} has brought unprecedented access to topological surface states. This has lead to several experimental breakthroughs including the observations of quantum Hall effects~\cite{QHE}, quantum anomalous Hall effect~\cite{QAHE1}, spin-polarized conduction of current~\cite{spin-to-charge2} and chiral Majorana modes~\cite{Majorana_TI_SC}. This progress has however come at a price. Compensation doping introduces strong electronic disorder that significantly deteriorates the quantum coherence of surface state carriers. Surface and bulk electron hole puddles~\cite{puddle3, puddle4, puddle5} have been shown to strongly couple to surface state electrons, leading to significantly enhanced dephasing rates compared to undoped TIs~\cite{banerjee2016,enhanced, xu2017disorder}. Such strong dephasing can mire future efforts at using topological insulators for information processing, quantum computing or spintronic applications that heavily rely on large coherence lengths of the surface states. It is therefore imperative to design topological insulators that posses high quantum coherence, while at the same time retaining surface dominated coherent transport.

In this work, we show that {\it structural} disorder rather than {\it compositional} disorder can be used to design TIs showing the above qualities. Our samples show strongly enhanced carrier coherence, while simultaneously suppressing the contribution of bulk currents to coherent transport, reflected as an electrical decoupling of the top and bottom topological surface states. Such decoupling has been previously achieved only through doping~\cite{cha2012weak, Lee2012, hsiung2013, decoupledSS, banerjee2016}. Surprisingly, we find that bulk structural disorder also alters the mechanism of carrier decoherence, that is usually left unaffected by static disorder. In a striking departure from the usual Nyquist type dephasing due to electron electron interactions where the phase coherence length $\lf \propto T^{-0.5}$, $T$ being the sample temperature, our samples in the thin limit show decoherence that is indicative of energy exchange due to direct frictional coupling between the opposite topological surface states with $\lf \propto T^{-1}$. 

\begin{figure}[!t]
\includegraphics[width=0.9\linewidth]{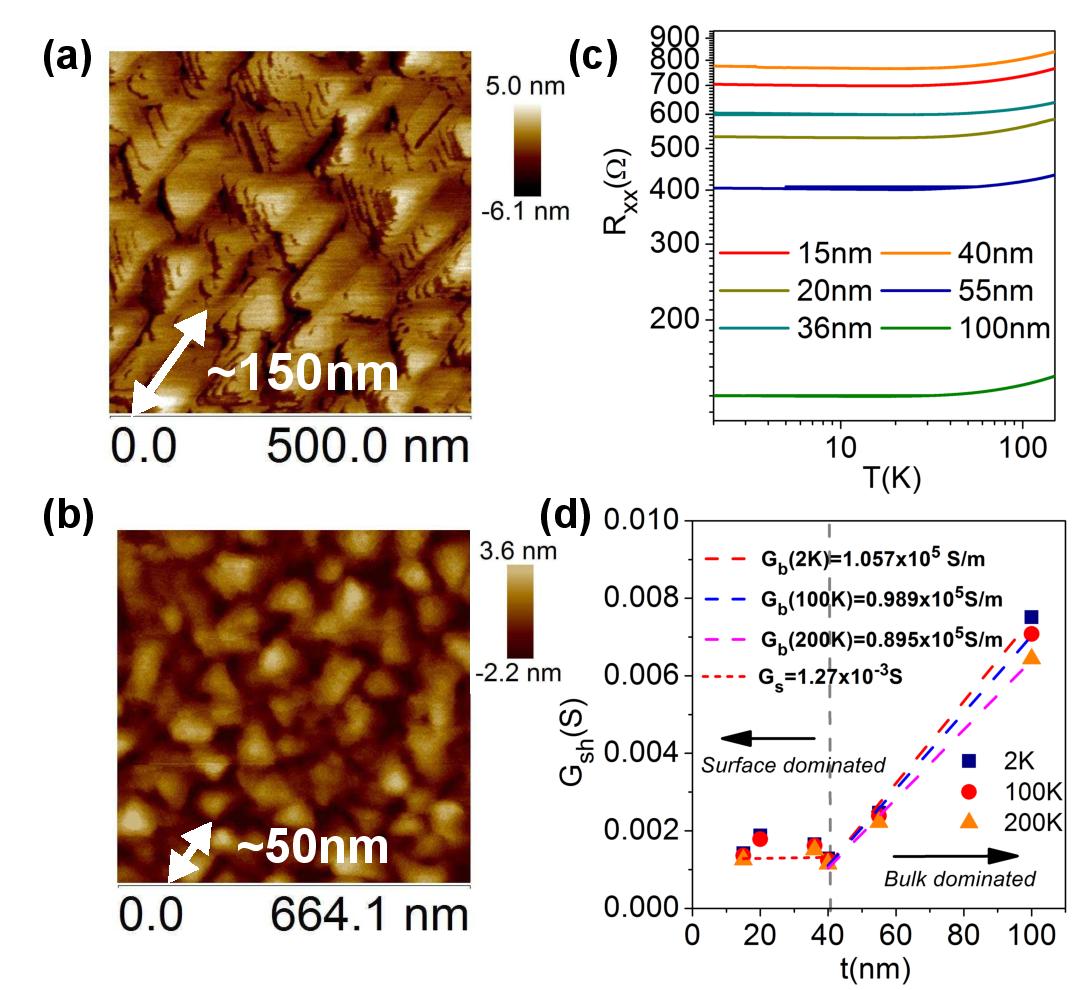}
\caption{(a) and (b) AFM images of samples with crystallite sizes of 150nm and 50nm respectively (c) Resistance vs temperature measurements for samples with different thicknesses (d) Sheet conductance G$_{sh}$ vs thickness}
\label{fig01}
\end{figure}

We postulate that structural disorder drastically affects the bulk conductivity, while leaving topological surface conductivity relatively unchanged, leading to our striking observations. To test this hypothesis, we perform magnetotransport in magnetic fields applied parallel to the sample plane that can be used to estimate the ratio of bulk to surface currents. This ratio shows a drammatic dependence on thickness and temperature, that indeed represents a strong suppresion of bulk coherent transport. Finally, by growing samples with different disorder strengths(but same thickness$\sim$40QL, 1QL $\simeq$ 1nm), we show that increased disorder indeed leads to surface-state decoupling, suppression of bulk currents and drastically improved carrier coherence.

\begin{figure}[!t]
\includegraphics[width=0.95\linewidth]{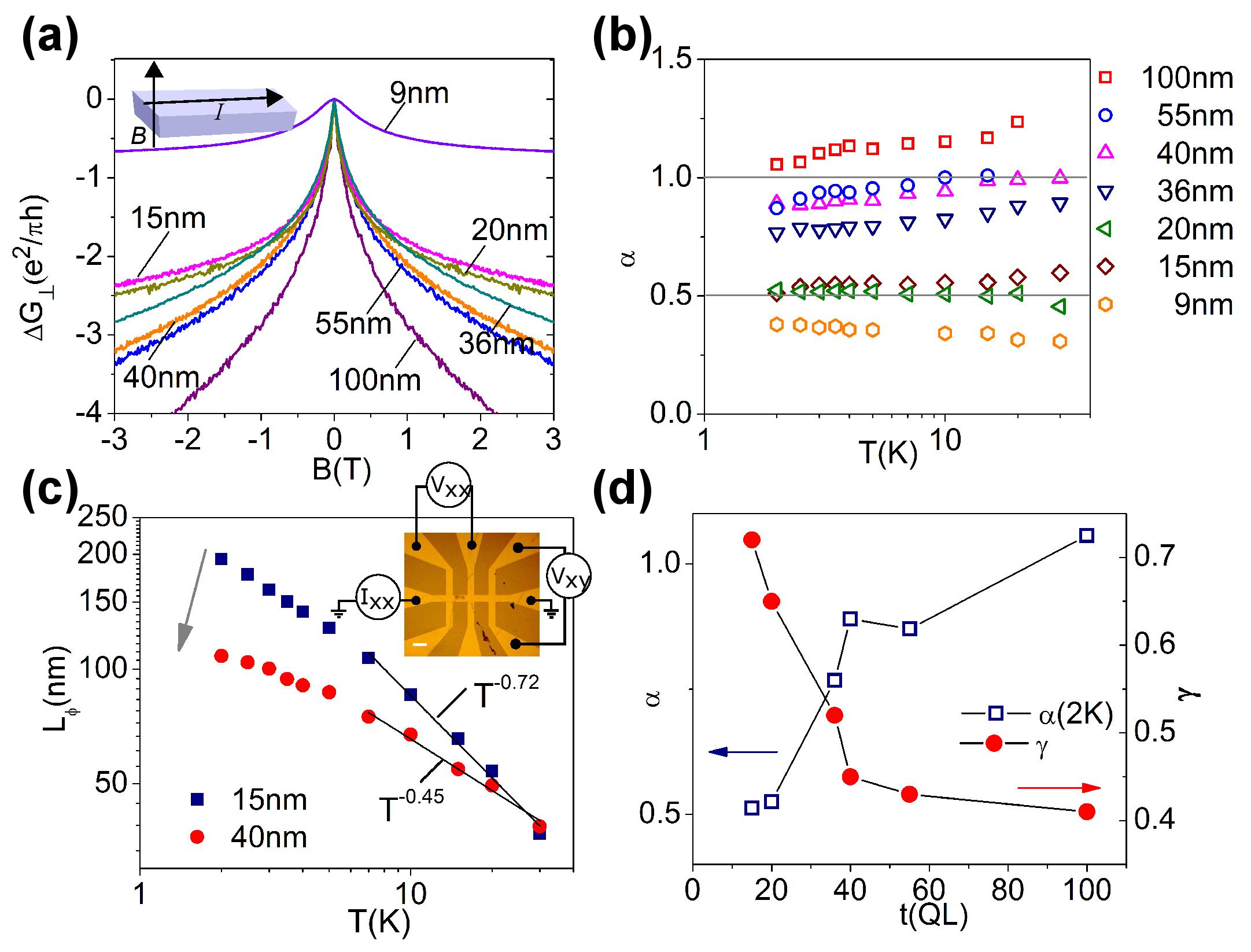}
\caption{(a) Perpendicular field magneconductance for samples with different thickness at $T=2K$. (b) $\alpha$ as a function of temperature (c) Phase coherence length (L$_{\phi} \propto T^{-\gamma}$) as a function of temperature for two different thicknesses. Inset: Device configuration. Scale=200$\mu$m (d) $\alpha$ and $\gamma$ as a function of sample thickness at T=2K.}
\label{fig02}
\end{figure}
 
Thin films of \BS are grown on Si(111) substrates using pulsed laser deposition, where we are able to controllably tune sample grain sizes as shown in Figs.~\ref{fig01}(a) and (b). All samples, regardless of grain size produce identical X-ray diffraction(XRD) patterns showing only (00L) peaks corresponding to \BS (see supplementary material section A). Details of thin film growth and characterization can be found in our previous works ~\cite{banerjee2017intermediate, banerjee2017granular} and supplementary material section A. Thin films are patterned into 8-probe Hall bars using optical lithography followed by Argon ion etching. Electrical transport measurements are performed down to 2K in magnetic fields up to 8~T.  

For the first part of the work, in all samples the lateral size of the individual \BS grains is kept constant at $\sim$ 50nm(Fig.~\ref{fig01}(b)). As we show later, this grain size provides an optimal degree of structural disorder that is crucial for our experiments. The bulk carrier density and mobility in all our samples lies in the range $n_{3D}=0.8-2 \times 10^{19}$/cm$^3$ and $\mu\sim$ 250 cm$^2$/V-s (T$=$2K) respectively resulting in a dimensionless conductivity of $g=k_F l_e \sim 6-10$; this is estimated from the Fermi momentum $k_F=(3 \pi^2 n_{3D})^{1/3} \sim 0.6$ ${\rm nm}^{-1}$ and mean free path $l_e=(\hbar \mu/e) k_F \sim 10$ nm. The dimensionless conductance $g$ lies in an intermediate regime, where although transport is diffusive($g>1$), the effect of disorder is still sizable. The large electronic carrier density generated by atomic point defects~\cite{xue2013first} ($n_{3D}\simeq10^{19}$/cm$^3$) provides strong electrostatic screening from charged dopants, making the disorder type primarily structural. This is verified by experiments, where we were able to significantly alter $g$ by changing grain size, while carrier density remained constant (supplementary material Tables S1 and S3). This is also consistent with previous experimental work where improving structural order was strongly correlated with large carrier mobility enhancements~\cite{topologicalprotection, thicknessindependentTI}. 

Resistance versus temperature(R-T) measurements shown in Fig.~\ref{fig01}(c) show linear metallic behavior in all samples. In Fig.~\ref{fig01}(d), we plot the sheet conductance $G_{sh}$ of samples at three different temperatures(2K, 100K and 200K) as a function of thickness. We observe a striking manifestation of surface dominated electrical transport, where $G_{sh}$ becomes independent of sample thickness for $t \leq 40$nm. Above this thickness conductance becomes bulk dominated with the $G_{sh} \propto G_b t$ . We extract a 3D bulk conductivity of $G_b \simeq 1.057 \times 10^5$ S/m at T=2K, and a 2D surface conductance of $G_s \simeq 1.27 \times 10^{-3}$S. 

To probe coherent transport in our samples, we use magnetoresistance measurements. Weak(anti)localization effects in magnetoresistance has emerged as a powerful tool to reveal coherent dynamics of charge carriers in a wide variety of diffusive systems, including thin films and crytalline flakes of topological insulators~\cite{chen2010gate,checkelsky2011bulk,chen2011tunable,steinberg2011electrically, cha2012weak,banerjee2017intermediate,topologicalprotection,powerlaw2,BSTS2,enhanced}. Fig.~\ref{fig02}(a) depicts the perpendicular field magnetoconductance(MC) data for samples with different thicknesses at $T=2K$. The MC data shows clear signatures of negative magnetoconductance due to the weak anti-localization(WAL) effect. In weakly disordered conductors with spin-orbit coupling, WAL gives rise to a quantum correction to conductance described by the Hikami-Larkin-Nagaoka (HLN) formula~\cite{HLN1,HLN3} as follows: $\Delta G_{xx}(B)=-\alpha \frac{e^2}{2\pi^2 \hbar} \left[\psi\left(\frac{1}{2} + \frac{\hbar}{4eL_\phi^2B}\right)-\ln \left(\frac{\hbar}{4eL_\phi^2B}\right)\right]$
in the limit of strong spin-orbit scattering. Here $\Delta G_{xx}$ is the change in sample 
conductivity, $B$ is the applied magnetic field, 
$L_\phi$ is the phase coherence length, and 
$\psi(x)$ denotes the digamma function. The leading constant $\alpha$ indicates the total number of {\it uncoupled} spin-orbit coupled channels, with a value of 0.5 for each such channel. 

As shown in Fig.~\ref{fig02}(b), $\alpha$ increases with increasing sample thickness, while its temperature dependence is rather weak. The transition from $\alpha=0.5$ to $\alpha=1.0$ is representative of a doubling of the number of coherent channels that contribute to electrical transport. In the thin film limit, the two topological surface states remain electrically coupled. Upon increasing the sample thickness, we observe that $\alpha$ transitions abruptly from 0.5 to 1 at $t\simeq$ 36nm(Fig.~\ref{fig02}(d)), indicating an electrical decoupling of the two topological surface states. While such decoupling has previously been observed in compensation doped TI samples with suppressed bulk carrier densities~\cite{cha2012weak, Lee2012, hsiung2013, banerjee2016}($\sim10^{16}$/cm$^3$), its observation in our pristine Bi$_2$Se$_3$ samples is surprising and demands a detailed discussion. At such large carrier concentrations($n_{3D}\simeq10^{19}$/cm$^3$), such a decoupling transition in unexpected. Also, at $T=2K$ the phase coherence length $L_{\phi}\sim110-200$~nm~$\gg t$ whereas the largest sample thickness is 100~nm. The decoupling of surface states at $t\sim$ 36~nm therefore cannot be due to inability of carriers to coherently couple the opposite surface states.

The decoherence mechanism of carriers also shows a striking departure from the usual Nyquist electron-electron(e-e) interaction observed rather universally in diffusive systems~\cite{rammer, altshuler1981, altshuler1982, altshuler1985}. Fig.~\ref{fig02}(c) depicts $\lf$ as a function of temperature for two samples. We evaluate the power law exponent $\gamma$, where $\lf \propto T^{-\gamma}$, and plot it in Fig.~\ref{fig02}(d) as a function of thickness. In the decoupled surface-state regime($\alpha=1$),  $\gamma \simeq 0.5$.   For two-dimensional metallic conductors, $\tau_\phi \propto \lf^2$ is usually proportional to $T^{-1}$ due to electron-electron(e-e) interaction from small-energy ($\epsilon<\hbar/\tau_e$) transfer~\cite{decoherencerate1}. This linear temperature corresponds to the so-called Nyquist dephasing regime and has been observed in TI samples by several groups~\cite{cha2012weak, kandala, powerlaw2, powerlaw3}. However, upon entering the coupled surface-state regime($\alpha=0.5$), $\gamma$ rises sharply and reaches values as large as 0.72. This represents a super-linear decoherence rate($2\gamma\simeq1.4$) and leads to significantly enhanced coherence at lower temperatures, as seem from the 15nm sample in Fig.~\ref{fig02}(c). As we show later, tuning the disorder slightly can lead to further suppression of decoherence rates with $\gamma=1$ ($\tau_\phi \propto T^{-2}$), and $\lf$ as high as $\sim$650nm at T=2K, representing a complete breakdown of the Nyquist mechanism. 

\begin{figure}[!t]
\includegraphics[width=0.95\linewidth]{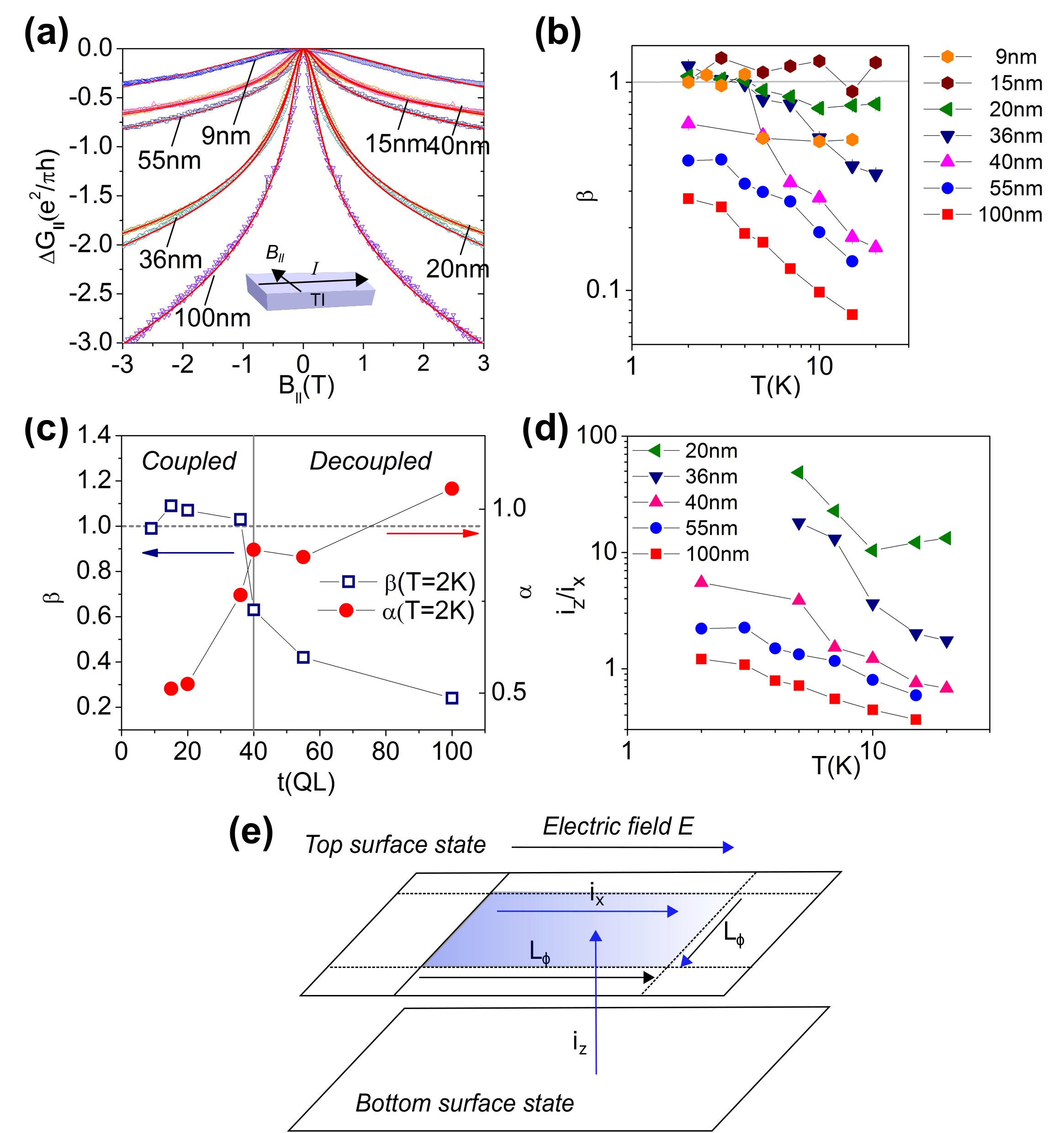}
\caption{(a) Parallel field magnetoconductance for different sample thicknesses at T=2K (b) Surface state coupling parameter $\beta$ obtained by fitting of the MC data (c) $\alpha$ and $\beta$ as a function of sample thickness at T=2K (d) $s=i_z/i_x$ as a function of temperature obtained from $\beta$ (e) (a) Schematic of mechanism of surface state coupling due to bulk currents $i_z$ coherent coupling opposite surface state. An opposite current flows from the top to bottom surface(not shown)}
\label{fig03}
\end{figure}

The unusual dependencies of surface state coupling parameter $\alpha$ and the decoherence rate $\gamma$ on sample thickness indicate a non-trivial role played by the bulk states in these experiments. To quantify bulk currents, we perform magnetoresistance measurements in magnetic fields applied parallel to the sample plane as shown in Fig.~\ref{fig03}(a). Electrons traversing the bulk of a sample pick up Aharanov-Bohm(AB) phases on completion of a loop, resulting in quantum interference. Application of a magnetic field parallel to the sample destroys this interference, resulting in a parallel field WAL effect~\cite{altshuler, dugaev, beenakker, sacksteder14} described by the following general formula: $\Delta G_{xx} (B_\parallel)=- \alpha \frac{e^2}{2\pi^2 
\hbar} \ln \left(1+\beta \frac{et^2}{4 \hbar B_{\phi}} B^2_\parallel \right)$,
where $B_\phi = \hbar/(4e\lf^2)$, $t$ is the sample thickness and $\beta$ is a measure of the bulk current that couples the opposite topological surface states. Fitting this equation to the parallel MR data shown in Fig.~\ref{fig03}(a) yields the coupling parameter $\beta$, with $\lf$ already known from perpendicular field measurements. As depicted in Figs.~\ref{fig03}(b),(c), $\beta$ shows a striking dependence on sample temperature and sample thickness. In the coupled SS regime, $\beta$ remains close to 1 (Fig. ~\ref{fig03}(c)), till around $t \sim 40$~QL, after which it degrades sharply at the onset of surface state decoupling. At T=2K, $\beta$ decreases $\sim$ 5 times as $t$ increases from 15~QL to 100~QL. More strikingly, the upward jump in $\alpha$ to 1.0 is simultaneous with the downward jump of $\beta$. For the thinnest samples ($t=$9nm,15nm,20nm), $\beta$ remains fairly constant with T. On the other hand, for thicker samples ($t=$36nm, 40nm, 55nm and 100nm), that lie in the decoupled SS regime($\alpha=1$), $\beta$ decays more and more strongly and follows a power law behavior as shown in Fig.~\ref{fig03}(b). Conventional models for $\beta$ predict no thickness or temperature dependence(supplementary material section D), therefore we resort to more general arguments.

For coherent coupling between two metallic layers, a transverse current $i_z$ must flow between them. The coupling strength is then decided by the ratio $i_z$ to the horizontal charge current $i_x$ flowing along the layer. $\beta$, measured from parallel field MR can be used to estimate this ratio $s=\frac{i_z}{i_x}$. The two layers get coupled($\alpha=0.5$) when $s \gg 1$ and decoupled ($\alpha=1.0$) when $s \ll 1$. From the Raichev-Vasilopoulos~\cite{RV}(RV) model that has been previously used to study TIs~\cite{lin}(see supplementary material sections D), $\beta$ depends on $s$ as follows:$\beta=2(1+s)/(1+2s)-ln(1+2s)/s.
\label{eqRV}$
We plot this ratio in Fig.~\ref{fig03}(d).  Within a phase coherent region of size $\lf$ on the sample surface (see Fig.~\ref{fig03}(e)) , the injected bulk current $i_z \propto \lf^2$, whereas the surface current $i_x$ remains independent of $\lf$.  The suppression of phase coherence with increasing temperature therefore explains the strong suppression of $i_z/i_x$. On the other hand, the strong dependence of $i_z/i_x$ on sample thickness $t$ represents enhanced scattering of bulk electrons due to longer path lengths $\propto t$. In the presence of scattering due to bulk disorder, $i_z$ can therefore be substantially suppressed with increasing thickness.

\begin{figure}[!t]
\includegraphics[width=1.0\linewidth]{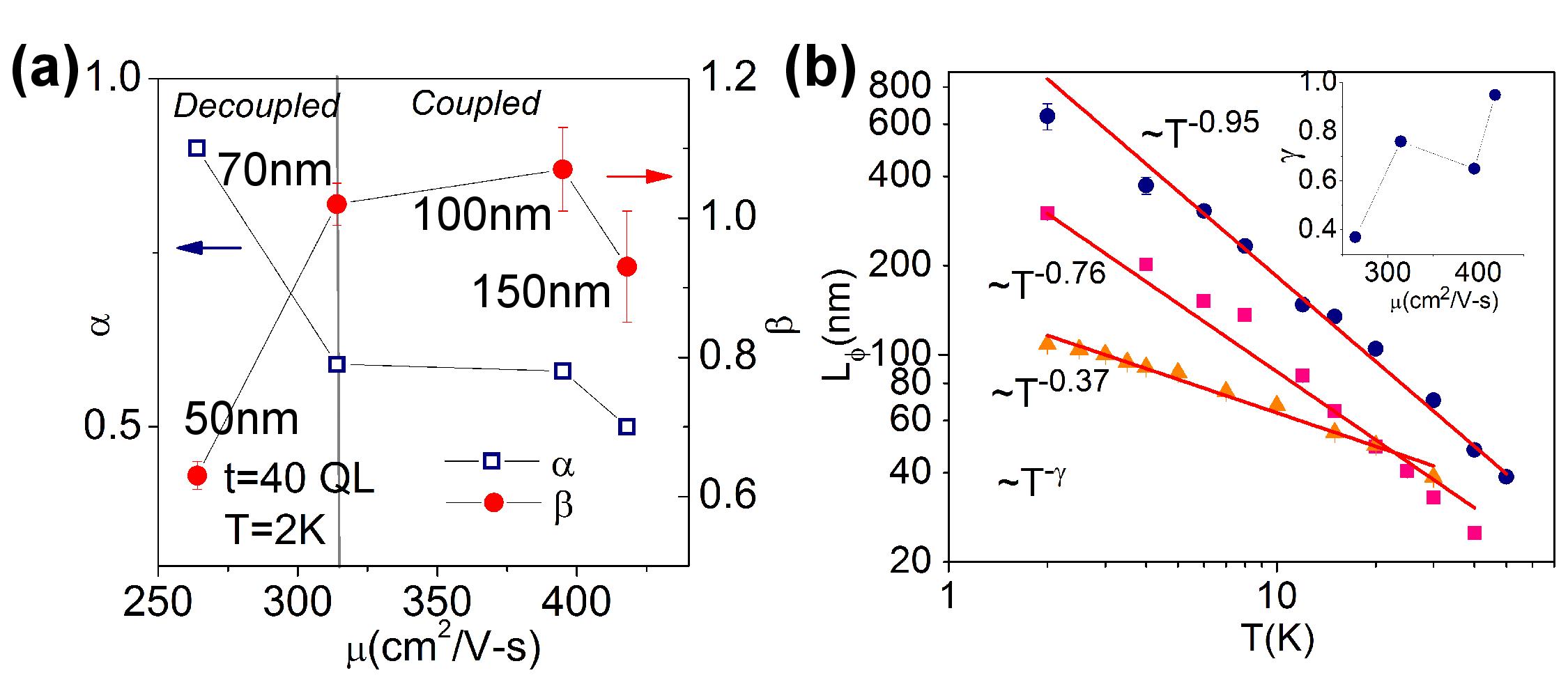}
\caption{(a) $\alpha$ and $\beta$ as a function of carrier mobility for 4 different samples with thickness of $\sim$ 40~QL and T$=$2K. Grain sizes of the four samples (50nm, 70nm, 100nm and 150nm) and are shown alongwith the corresponding values of $\beta$ and $\mu$ (b) $\lf$ as a function of temperature for samples with different disorder. Inset: $\gamma$ as a function of sample mobility $\mu$. }
\label{fig04}
\end{figure}
 
To further verify the suppression of bulk currents by disorder, we perform a disorder dependent study. We grow thin film samples of \BS with roughly the same thickness $t \sim 40$~QL but with different levels of structural disorder in the form of grain boundaries modified by varying grain sizes in the range $\sim$50nm to $\sim$150nm(Fig.~\ref{fig01}(a),(b)). We chose this thickness because it appears to correspond to a critical length required for surface state decoupling. The variation in elastic disorder strength is quantified in terms of the Hall carrier mobility($\mu$) extracted at T=2K, that varies from 250 cm$^2$/V-s to 420 cm$^2$/V-s. Fig.~\ref{fig04}(a) shows the variation of $\alpha$ and $\beta$ as a function of $\mu$. As $\mu$ increases, the surface states become progressively more coupled with $\alpha$ varying from $\sim$1.0 to $\sim$0.5. Simultaneously, $\beta$ rises sharply indicating an increase in bulk currents, confirming our hypothesis. It must be noted that a variation of bulk structural disorder produces a concomitant variation of surface disorder that can change surface transport properties, apart from bulk transport properties that we focus on. However, this effect does not significantly change our interpretation of results as we discuss in supplementary material section F.

More significantly, as the disorder strength is decreased, the electronic dephasing rate changes drastically from $\lf \propto T^{-0.37}$ to $\lf \propto T^{-0.95}$ as shown in Fig.~\ref{fig04}(b). As the bulk mobility increases, $\gamma \to 1$ as shown in the inset of Fig.~\ref{fig04}(b). This leads to strongly enhanced carrier coherence at low temperatures as seen from Fig.~\ref{fig04}(b): at large sample temperatures(T$\sim$20-30K), all samples exhibit similar values of $\lf \simeq 30-50$nm. However, when sample temperature is lowered, carrier decohernce with $\lf \propto T^{-0.5}$ leads to smaller values of $\lf$ at low temperatures, compared to $\lf \propto T^{-1}$. Such decoherence rates are not possible in two dimensional systems, where $\lf \propto T^{-0.5}$ ($\tau^{-1}_\phi \propto T$) is universally observed in experiments~\cite{rammer} and appears naturally due to the fluctuation-dissipation theorem, with the spectral power of electromagnetic fluctuations $\propto T$.


A natural explanation to this unconventional dephasing mechanism can be found by considering frictional Coulomb drag(CD) between the opposite topological surface states~\cite{InteractionSHE3, drag2, drag3}. In bilayer metallic systems that are weakly tunnel coupled, direct electron-electron interaction between the two layers is known to be the dominant scattering mechanism. This has been observed in several experiments on bilayers of two-dimensional electron gas systems including graphene~\cite{InteractionSHE3, drag2, drag3, drag1, drag_graphene} where $\tau_{ee,dr}^{-1} \propto T^{2}$, the factor of 2 arising from scattering between two different Fermi surfaces. The dephasing length therefore follows $\lf \propto T^{-1}$, exactly as observed in our experiments. In this regime $\tau^{-1}_{ee} \propto t^{-4}$, $t$ being the separation between the two layers, which explains why we observe this effect only in the thin film limit. While different independent dephasing processes affect the coherence length, the mechanism with the largest scattering dominates. We can therefore describe the dephasing process in our system as combination of CD and surface-bulk(SB) e-e interactions: $\tau^{-1}_\phi \propto \tau^{-1}_{ee,cd}+\tau^{-1}_{ee,sb} \simeq C_1 T^2/t^4 + C_2 T$. For small thicknesses, the CD mechanism dominates, while for larger thicknesses SB scattering becomes dominant and thereby restores the usually observed Nyquist mechanism(Fig.~\ref{fig02}(c)). Presumably, strong surface-bulk scattering in previous experiments on TIs have prevented an observation of the Coulomb drag mechanism. 

We now provide a coherent picture to explain all our experimental observations. The bulk carriers execute quantum diffusion and electrically connect the top and bottom surface states, while at the same time interacting with surface electrons through electron-electron scattering. For a sample with thickness $t$, the timescale for a bulk electron to traverse the thickness of the sample is $\tau_b\simeq t^2/D_b$, where $D_b$ is the bulk diffusion constant. For a bulk electron to coherently couple the two surface states, we require $\tau_b < \tau_{\phi}$, which is clearly not satisfied when $D_b$ is suppressed. The variation of surface-bulk e-e scattering rate with the bulk diffusion constant is however more complicated and will depend on both the bulk($D_b$) and surface($D_s$) diffusion constants, and has not been evaluated theoretically so far. Qualitatively however, it is known that $\tau^{-1}_{ee} \propto \log(D \nu)/D\nu$ in two dimensions~\cite{rammer, altshuler1982}, where $\nu$ is the bulk density of states. For strongly suppressed bulk diffusivities we have $D \nu \to 1$, implying a proximity to bulk state localization in which case $\tau^{-1}_{ee,sb} \to 0$. Clearly, this regime of dephasing in TIs merits more theoretical attention. Additionally, it is now understood that strong bulk disorder can completely localize bulk states, while failing to localize topological surface states that are inherently protected against Anderson localization~\cite{TAI,guo2010,agarwala2017topological}. Such a scenario can lead to fully surface dominated topological insulators, despite the large density of (localized) bulk carriers and has been verified in recent experiments~\cite{du2016,TIAndersonlocalization,banerjee2017granular}. 

In conclusion, our work provides unprecedented insights into carrier decoherence in topological insulators. We show that the usual Nyquist dephasing mechanism reported in TIs is an effect of strong bulk-surface electron-electron interaction. In the coupled surface state regime, upon suppressing bulk carrier diffusion by introducing structural disorder in our samples, we subdue bulk-surface e-e interactions thereby uncovering a hitherto unobserved dephasing regime dominated by energy loss due to frictional drag between the two topological surface states. Adding structural disorder also leads to the electrical decoupling of the two topological surface states and surface dominated electrical transport, which has never been observed for bulk conducting TIs. Our approach of using structural disorder rather than compositional disorder to achieve surface dominated conduction and enhanced carrier coherence solves one of the most pressing problems surrounding this field.    

{\bf Supplementary Material}
See supplementary material for details of sample preparation, additional electrical transport data and data analysis.

{\bf Acknowledgements}
A.B. thanks MHRD, India and P.S.A.K. acknowledges Nanomission, DST, India.


%

\end{document}